\documentclass{vgtc}                          

\graphicspath{{figures/}{pictures/}{images/}{./}} 

\usepackage{times}                     

\usepackage{tabu}                      
\usepackage{booktabs}                  
\usepackage{lipsum}                    
\usepackage{mwe}                       

\usepackage{mathptmx}                  

\usepackage{subcaption}
\usepackage{enumitem}

\onlineid{0}

\vgtccategory{Research}

\vgtcinsertpkg

\title{Rethinking Privacy Indicators in Extended Reality: Multimodal Design for Situationally Impaired Bystanders
}

\author{Syed Ibrahim Mustafa Shah Bukhari\thanks{e-mail: simsb@vt.edu} %
\and Maha Sajid\thanks{e-mail:mahas@vt.edu} %
\and Bo Ji\thanks{e-mail:boji@vt.edu}
\and Brendan David-John\thanks{e-mail:bmdj@vt.edu}}
\affiliation{\scriptsize Virginia Tech}

\teaser{
  \centering
  \begin{minipage}[b]{0.37\textwidth}
    \includegraphics[width=\linewidth]{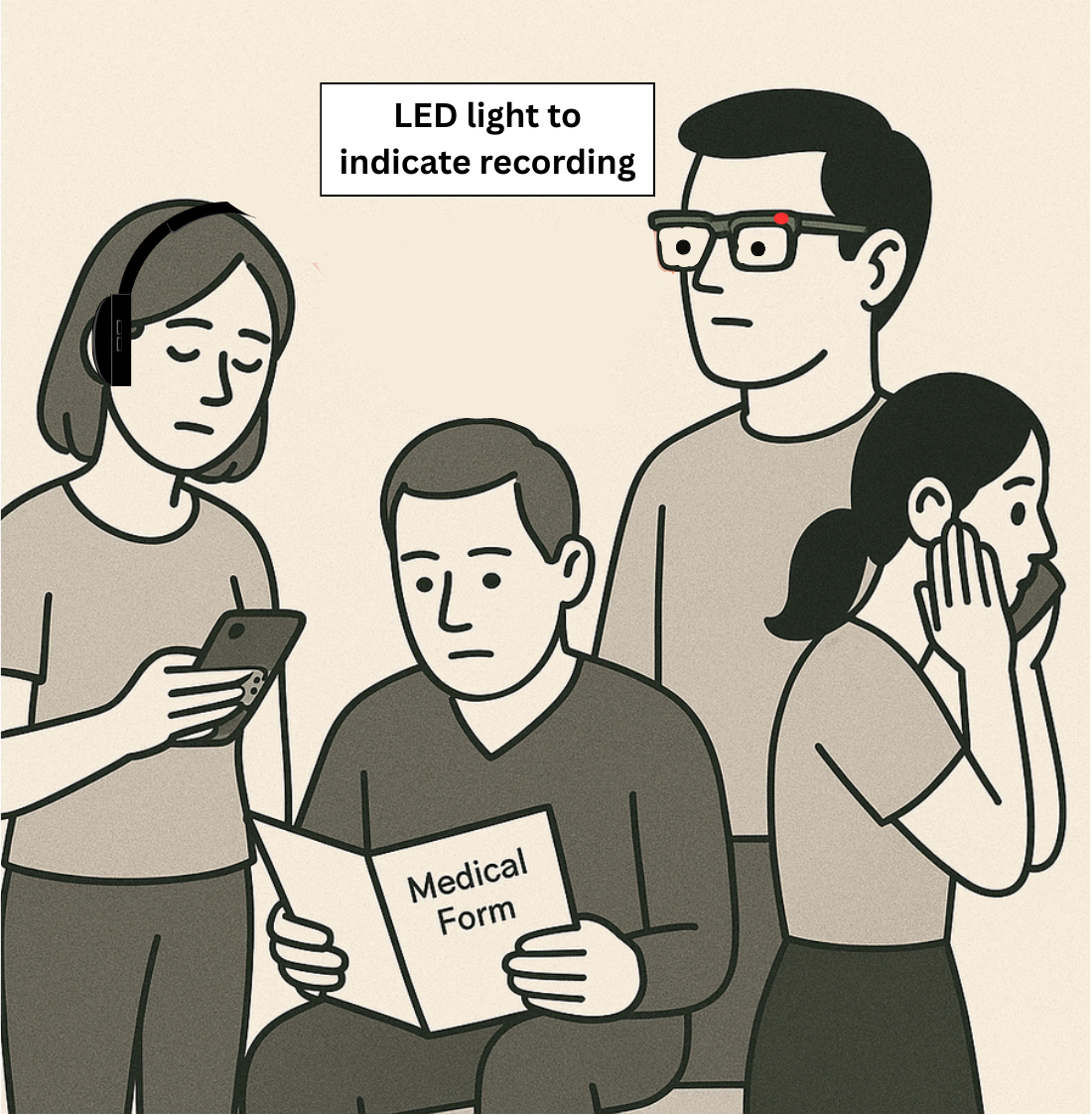}
    \caption*{(a)}
  \end{minipage}
  \begin{minipage}[b]{0.393\textwidth}
    \includegraphics[width=\linewidth]{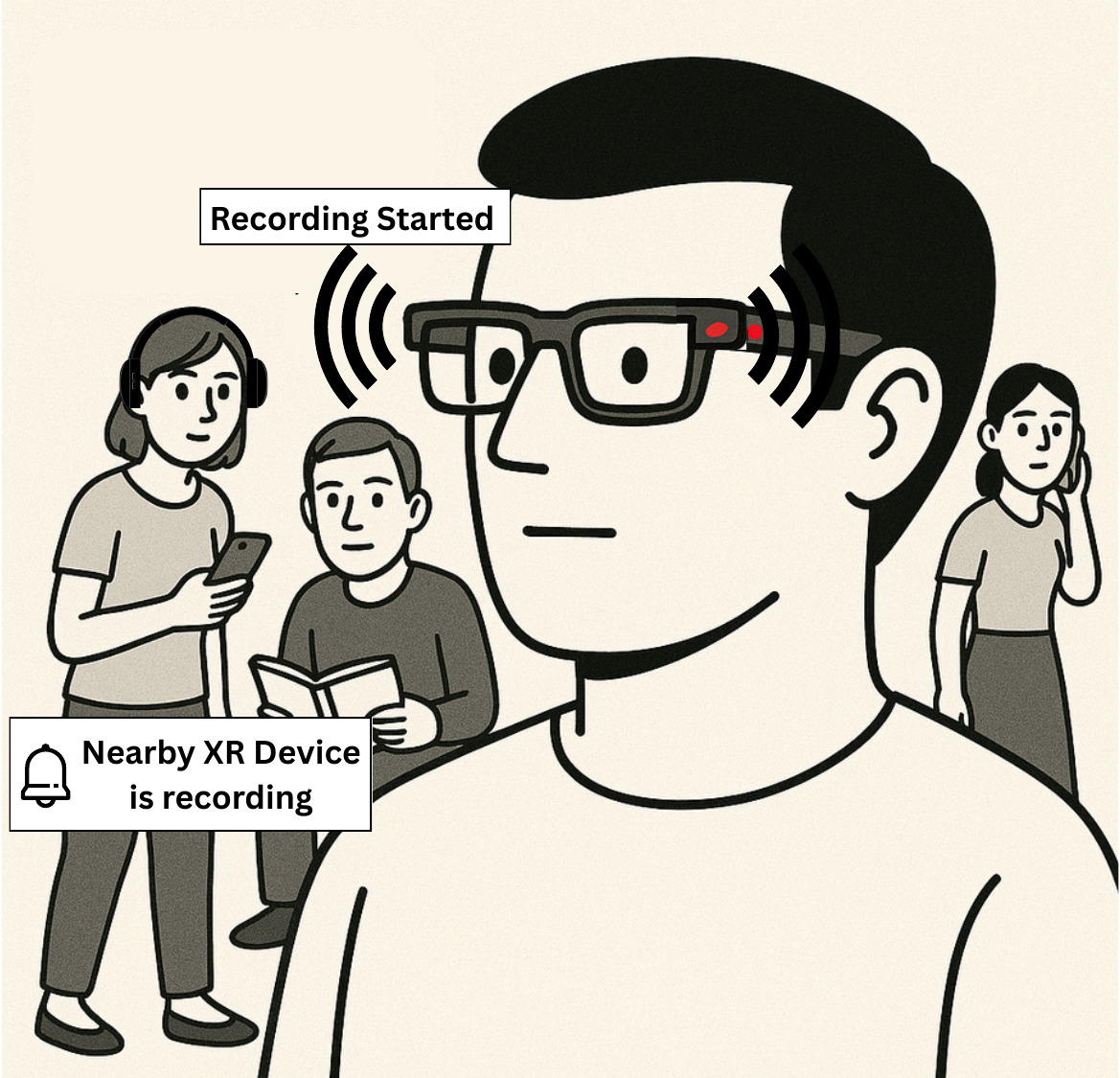}
    \caption*{(b)}
  \end{minipage}
  \caption{(a) Current XR privacy indicators, such as outward-facing LEDs, often go unnoticed by distracted or situationally impaired bystanders. As a result, private information, such as documents, phone screens, or personal conversations, may be recorded without their awareness. (b) Multimodal privacy indicators, such as combining a visible LED, an audio cue from the XR device, and a notification on the bystander’s phone, are more likely to reach situationally impaired bystanders who may be looking elsewhere or wearing headphones.
  }
  \label{fig:teaser}
}

\abstract{
As Extended Reality (XR) devices become increasingly prevalent in everyday settings, they raise significant privacy concerns for bystanders: individuals in the vicinity of an XR device during its use, whom the device sensors may accidentally capture. Current privacy indicators, such as small LEDs, often presume that bystanders are attentive enough to interpret the privacy signals. However, these cues can be easily overlooked when bystanders are distracted or have limited vision. We define such individuals as \textit{situationally impaired bystanders}. This study explores XR privacy indicator designs that are effective for situationally impaired bystanders. A focus group with eight participants was conducted to design five novel privacy indicators. We evaluated these designs through a user study with seven additional participants. Our results show that visual-only indicators, typical in commercial XR devices, received low ratings for perceived usefulness in impairment scenarios. In contrast, multimodal indicators were preferred in privacy-sensitive scenarios with situationally impaired bystanders. Ultimately, our results highlight the need to move toward adaptable, multimodal, and situationally aware designs that effectively support bystander privacy in everyday XR environments.
} 


\begin{document}


\firstsection{Introduction}

\maketitle

Modern Extended Reality (XR) devices span a range of form factors, from conspicuous headsets like Meta's Quest headsets~\cite{VRcompare} and Microsoft's HoloLens~\cite{Joyjaz}, to subtle smart glasses such as Meta Ray-Bans~\cite {RayBanMeta} and Snap Spectacles~\cite{SnapSpectacles}. As XR devices become increasingly integrated into everyday life~\cite{ADRIANACARDENASROBLEDO2022101863}, they pose privacy risks to bystanders: individuals in the vicinity of an XR device, whom the device sensors may accidentally capture during its use.

XR devices are equipped with a plethora of sensors, including outward-facing cameras, microphones, and depth sensors, that continuously capture information from the user's surroundings. While essential for enabling immersive user experiences such as gesture recognition and voice-based interaction, these sensors indiscriminately collect data from bystanders, often without their knowledge or consent~\cite{10339660}. Bystanders lack any meaningful mechanism to control or opt out of this data collection. Prior work has shown that such data can be used to infer sensitive personally identifiable information (PII), including facial identity, sex, race, age, and sexual preference~\cite{kroger2020does, liebling2014privacy, wenzlaff2016video, renaud2002measuring, borras2012depth, cheng20173d}, posing significant privacy risks to bystanders. Hence, there is a pressing need to develop mechanisms that inform bystanders of such ongoing data collection.

Many XR devices use privacy indicators to signal when sensing is active. For example, Magic Leap~2 features a small front-facing LED light that activates when outward-facing sensors are in use~\cite{MagicLeapLEDGuide}. Similarly, Meta Ray-Bans has a ``Capture LED'' to indicate when the device is recording~\cite{MetaPrivacyPage}. However, these privacy indicators rely on the visual attention of bystanders, limiting their effectiveness when bystanders are distracted, facing away, or have limited vision. We define such individuals as \emph{situationally impaired bystanders}: bystanders whose ability to access and perceive privacy indicators is hindered by temporary or contextual factors. Existing XR privacy indicators often rely on generic visual or auditory cues, assuming full attention and unimpaired perception. Moreover, such indicators can be easily concealed~\cite{Ashworth_2025}, further increasing the likelihood that situationally impaired bystanders remain unaware of ongoing data collection and potential privacy breaches. In this paper, we identify the lack of inclusive privacy indicators for situationally impaired bystanders and explore the design space for these indicators. We also evaluate some novel XR privacy indicator designs.

To this end, we conducted a focus group with eight participants (four pairs), which resulted in five novel XR privacy indicator designs for situationally impaired bystanders. We then evaluated these designs for their usability and perceived effectiveness with a user study of seven additional participants. The focus group and subsequent user studies were approved by Virginia Tech’s Institutional Review Board. Our main contributions are as follows:

\begin{itemize}
    \item {We identify and explore a previously overlooked gap in the design of XR privacy indicators by focusing on the needs of situationally impaired bystanders.}
    \item {We propose five novel privacy indicators designed to accommodate situational impairments, and evaluate their usability through a user study, finding that all meet or exceed the usability threshold of System Usability Scale (SUS).}


    \item {We offer design insights for the development of future privacy indicators that are inclusive of real-world bystander contexts, highlighting the need for adaptable, situationally aware, and multimodal (e.g., audio, visual, and haptic) designs that remain effective for bystanders with limited attention.}
    
\end{itemize}

\section{Related Work}\label{related-work}

The following subsections review prior work on existing implementations of bystander-facing privacy indicators~\ref{relatedwork-privacyindicators}, the role of situational impairment in shaping inclusive design practices~\ref{relatedwork-siandinclusivedesign}, and design approaches for privacy in emerging technologies~\ref{relatedwork-codesign}.

\subsection{Privacy Indicators for Bystanders}\label{relatedwork-privacyindicators}
 
Previous studies have found that bystanders are often uncomfortable around XR devices due to fear of being recorded unknowingly~\cite{denning2014situ}. Although there is ongoing research into the development of Privacy-Enhancing Technologies (PETs) to protect bystander privacy, current privacy protection mechanisms in XR remain insufficient. Some systems place control in the hands of the wearers, which limits the ability of bystanders to know whether privacy protection mechanisms are actively operating on the device or not~\cite{steil2019privaceye, corbett2023bystandar}. Whereas, other systems require bystanders to explicitly declare their privacy preferences~\cite{li2016privacycamera, koelle2018your}. However, this approach is only feasible if bystanders can recognize the presence of potential privacy risks in their environment. These challenges highlight the need for explicit privacy indicators to inform bystanders, particularly those who are situationally impaired.

Researchers have explored how XR devices can signal to nearby bystanders when sensors such as cameras are active. Koelle et al. found that traditional LED status lights are often too subtle to effectively communicate recording activity. In their study, participants expressed a strong preference for physical camera shutters over LEDs, as the visible occlusion provided greater assurance that no video was being captured~\cite{koelle2018beyond}. Similarly, Bhardwaj et al. reported that the sleek and minimalist design of modern AR glasses often conceals the presence of cameras, making it difficult for bystanders to recognize when recording is taking place~\cite{bhardwaj2024focus}. Some researchers have proposed richer and more context-aware signaling mechanisms. For instance, O’Hagan et al. found that bystanders’ preferences for notifications vary depending on the type of data being captured and their relationship to the wearer~\cite{o2023privacy}. Although simple status lights may suffice for routine tasks, more transparent cues are preferred in privacy-sensitive scenarios. However, for any indicator to be effective, it must be both noticeable and intuitive. These challenges highlight the need for further refinement of privacy indicators, ensuring that they are not only effective in design but also adaptable to varying real-world conditions.

\subsection{Situational Impairment and Inclusive Design}\label{relatedwork-siandinclusivedesign}

Recent work has recognized the impact of situationally induced impairments (SIIDs) on user interaction within XR environments. Liu et al. propose Human I/O, a unified framework for detecting SIIDs by estimating the availability of human input/output channels using egocentric vision, multimodal sensing, and large language models~\cite{liu2024human}. While their work effectively addresses how XR systems can adapt to users experiencing distractions, poor lighting, or multitasking, its primary focus remains on optimizing interaction for the situationally impaired user of the XR device. In contrast, our work extends the concept of situational impairment to bystanders—individuals in the shared physical environment who may be distracted or cognitively limited temporarily and thus unable to perceive traditional privacy cues. This shift highlights the need for privacy indicators that are inclusive not only of users' conditions but also of the situational contexts faced by the bystanders.

\begin{table*}[t]
    \centering
    \renewcommand{\arraystretch}{1.25}  
    \begin{tabular}{|p{0.28\textwidth}|p{0.65\textwidth}|}
        \hline
        \textbf{Privacy Indicator (PI)} & \textbf{Description} \\
        \hline
        Visual-Static (PI\textsubscript{VS}) & Imagine an XR device with a large, outward-facing screen that spans the front of the device. Whenever the user begins recording, a bright, neon-style message appears on the screen to clearly indicate the activity (e.g., ``Recording in Progress''), so bystanders are aware that they are being recorded. \\
        \hline
        Visual-Dynamic (PI\textsubscript{VD}) & Imagine an XR device with a large, outward-facing screen that spans the front of the device. The screen shows a live feed of what the user is currently recording. Additionally, a side panel displays the faces that have been captured by the device in the past minute, making bystanders aware if they were recorded. \\
        \hline
        Proximity-Multimodal (PI\textsubscript{PM}) & Imagine that XR devices are linked to nearby smartphones. When an XR device begins recording, all smartphones in the vicinity receive a notification (with audio, visual, and haptic cues) informing bystanders that a nearby XR device is actively recording. Each bystander can customize how this notification appears on their device. \\
        \hline
        Gaze-Multimodal (PI\textsubscript{GM}) & Imagine an XR device that tracks wearers' eye gaze. If the wearer looks directly at a bystander while sensors are actively recording, that bystander’s smartphone receives a notification, alerting them that they are being observed during recording. \\
        \hline
        Activity-Multimodal (PI\textsubscript{AM}) & Imagine XR devices with outward-facing indicators that adapt to the level of recording activity. When the camera is on (not recording), a small light appears. During active recording, a larger blinking light and a repeating tone are triggered. If a bystander is detected during recording, the light flashes rapidly, and the tone's volume increases to draw attention. \\
        \hline
    \end{tabular}
    \caption{Novel XR privacy indicator designs proposed for situationally impaired bystanders by focus group participants.}
    \label{tab:proposed-privacy-indicators}
\end{table*}

Existing XR privacy indicators often assume that bystanders are fully attentive and have unimpaired perception. Therefore, these privacy indicators rely primarily on visual cues to signal when sensors like cameras or microphones are active. However, such cues can be ineffective for situationally impaired bystanders. Zhao et al. highlighted the challenges visually impaired individuals face in recognizing recording devices and emphasized the need for non-visual indicators such as spatialized audio or haptic feedback to support equitable awareness~\cite{zhao2023if}. Similarly, Ahmed et al. developed wearable systems that combined audio and tactile feedback to convey situational information, demonstrating improved awareness for users with reduced sensory access~\cite{ahmed2019conveying}. While these works show efforts toward inclusive design strategies, they only focus on a specific set of impairments and do not address the broader category of situationally impaired bystanders. This gap highlights the need for privacy indicators that remain effective in varying situational constraints and sensory conditions.

\subsection{Designing for Privacy in Emerging Technologies}\label{relatedwork-codesign}

Our study methodology follows an iterative design process, allowing for the refinement of privacy indicator systems by combining expert-driven design with user feedback. This approach aligns with design methodologies in prior studies on emerging technologies.

Paneva et al. conducted brainstorming and sketching sessions with novice game developers and designers, followed by privacy expert evaluations, to explore and refine privacy interfaces tailored for Virtual Reality (VR)~\cite{paneva2025usable}. Their study identified key challenges such as balancing user engagement with privacy awareness and managing complex privacy information with user comprehension. These findings informed our study design, particularly structuring focus group sessions to generate various novel concepts of XR privacy indicators for situationally impaired bystanders.

Similarly, Rajaram et al. conducted an elicitation study involving both Augmented Reality (AR) experts and Security and Privacy (S\&P) experts to develop privacy-informed sharing techniques for multi-user AR settings~\cite{rajaram2023eliciting}. Their approach emphasized the importance of integrating S\&P considerations into the design process through expert collaboration. This aligns with our methodology of engaging domain experts to generate novel privacy indicator designs for situationally impaired bystanders in XR environments.

We grounded our study design in these methodologies to ensure that our approach is informed by established practices in the field, aiming to develop effective and user-centered XR privacy indicators for situationally impaired bystanders. 

\section{Focus Group}\label{focusgroup}

We conducted focus group sessions with domain researchers to generate novel concepts of XR privacy indicators for situationally impaired bystanders. The following question guided these sessions:

\textbf{RQ1:} \textit{How can privacy indicators be designed to accommodate the needs of situationally impaired bystanders in XR environments?}

\subsection{Methodology}\label{focusgroup-methodology}
\subsubsection{Participants}
We recruited participants until thematic saturation was met, i.e., when additional focus group discussions no longer produced new themes or insights relevant to the study. This process resulted in eight participants (three females) and four focus group sessions, each with two participants. All participants were familiar with the privacy and security concerns of emerging technologies. The most experienced participants were active XR users and researchers, while the least experienced had some experience using XR devices and some knowledge of privacy risks for bystanders. We chose not to recruit from the general population, as the focus group required a prior understanding of XR technology and its privacy implications.

\subsubsection{Procedures}
Participants received an introductory email and the consent form at the time of recruitment. The introductory email allowed participants to familiarize themselves with the concept of privacy indicators in XR systems and situational impairment. Each focus group session began with a primer presentation, covering key concepts such as bystander privacy issues, existing privacy indicator systems in XR, and situational impairment. After the presentation, the participants asked clarification questions if needed.

The participants were then given 30 minutes to brainstorm and sketch ideas together for XR privacy indicator systems for situationally impaired bystanders. At the end of the session, each pair presented their ideas, explaining the functionality of their designs, the rationale behind their choices, and how their systems would address the needs of situationally impaired bystanders. The presentations were independently analyzed and coded by researchers. The researchers collaboratively reviewed and discussed results to identify common themes and consolidate unique privacy indicators. 

\subsection{Results}\label{focusgroup-results}

The focus group sessions yielded a total of seven candidate XR privacy indicator concepts designed to accommodate situationally impaired bystanders. To refine this set, we analyzed the candidate indicators to identify any functional overlaps in their modality and triggering mechanism. This allowed us to consolidate the candidate indicators into five distinct ones, each offering a unique combination of sensory modality and activation conditions.

Table~\ref{tab:proposed-privacy-indicators} presents the five consolidated privacy indicator designs produced from our focus group. The descriptions shown in the table are exactly as they were presented to participants in the subsequent user study. The final set includes a mix of visual and multimodal approaches, varying in modality, activation mechanism, and contextual awareness. PI\textsubscript{VS} and PI\textsubscript{VD} rely on outward-facing displays that either show static alerts or real-time visualizations of the recording context. PI\textsubscript{PM} and PI\textsubscript{GM} use smartphone-based notifications triggered by proximity and gaze, respectively, enabling awareness even when visual attention is limited. PI\textsubscript{AM} uses escalating visual and auditory cues that intensify based on the level of recording activity and the presence of bystanders. Together, these indicators represent a diverse design space for inclusive, context-sensitive bystander privacy indicators for XR environments.

\section{User Study}\label{userstudy}

We then conducted a survey to evaluate the usability, contextual effectiveness, and user preferences of each proposed indicator system. The study was guided by the following question:

\textbf{RQ2: }\textit{How do usability and context influence the perceived effectiveness of different privacy indicators for bystanders under situational impairment?}

\subsection{Methodology}\label{userstudy-methodology}

\subsubsection{Participants}

We performed an a priori power analysis, which indicated that 28 participants would be required to achieve the desired statistical power. However, due to the exploratory nature of this work and resource constraints, we recruited a new set of seven participants (two females), including four undergraduate students and three graduate students. The participants were recruited through the Computer Science graduate student listserv and the university-wide Center for Human-Computer Interaction (CHCI) listserv at Virginia Tech. Participants without prior experience with XR devices were excluded, as a basic understanding of XR technology was required to evaluate privacy indicator systems effectively. Participants' experience levels varied, with three being somewhat familiar with XR devices, two being familiar, and two being very familiar.

\subsubsection{Procedure}
During the user study, each participant was assigned a random participant ID to ensure anonymity. Participants were also provided with a consent form to read and sign.

At the start of each session, participants saw a primer presentation that introduced the concept of privacy indicators in XR systems. Participants were given time to ask questions and seek clarification on any aspects of the study.

Following the presentation, participants were asked to complete a survey questionnaire. The survey was divided into six sections. The first five sections focused on individual bystander privacy indicator systems, with each section corresponding to a system proposed in the focus group. Participants first received an overview of each system in form of descriptions, as shown in Table~\ref{tab:proposed-privacy-indicators}, followed by the opportunity to ask questions or request clarifications. They were then asked to rate the usefulness of the system within a specific scenario and respond to a set of System Usability Scale (SUS) questions to evaluate the system’s usability~\cite{brooke1996sus}.

In the final section, participants were asked to review all the privacy indicator systems and provide an overall ranking, along with their reasoning for the rankings. This allowed for a comprehensive assessment of the participants’ preferences and insights into the effectiveness of each system.

\begin{table}[t!]
    \centering
    \renewcommand{\arraystretch}{1.25}
    \begin{tabular}{|l|p{0.5\linewidth}|}
        \hline
        \textbf{Scenario} & \textbf{Description} \\
        \hline
        Classroom Waiting (S\textsubscript{CW}) & In a classroom, waiting for your professor, a classmate is recording with their XR headset. \\
        \hline
        Classroom Exam (S\textsubscript{CE}) & In a classroom, taking an open-book exam, a classmate is recording with their XR  headset. \\
        \hline
        Park Bench (S\textsubscript{PB}) & In a park, sitting on a bench, a person nearby is recording with their XR headset. \\
        \hline
        Park Reading (S\textsubscript{PR}) & In a park, reading a book, a person nearby is recording with their XR headset. \\
        \hline
        Bank Waiting (S\textsubscript{BW}) & In a bank, waiting in line, a person beside you is recording with their XR headset. \\
        \hline
        Bank Form (S\textsubscript{BF}) & In a bank, filling out personal information, a person beside you is recording with their XR headset. \\
        \hline
        Beach Sunbathing (S\textsubscript{BS}) & On a beach, sunbathing, a person beside you is recording with their XR headset. \\
        \hline
        Beach Volleyball (S\textsubscript{BV}) & On a beach, playing volleyball, a person watching the match is recording with their XR headset. \\
        \hline
    \end{tabular}
    \caption{Privacy Scenarios}
    \label{scenario_table}
\end{table}

\textbf{Scenarios: }\label{userstudy-methodology-scenarios}In the scenario-specific questions, participants were asked to imagine themselves as bystanders and rate the usefulness of each privacy indicator system on a five-point scale, ranging from ``not very useful'' to ``very useful''. Participants rated each system based on how effectively it alerted them, as bystanders, to potential privacy risks, both in the presence and absence of privacy-sensitive conditions and situational impairments.

A total of eight scenarios were presented, as shown in Table~\ref{scenario_table}. These scenarios included four indoor (S\textsubscript{CW}, S\textsubscript{CE}, S\textsubscript{BW}, S\textsubscript{BF}) and four outdoor (S\textsubscript{PB}, S\textsubscript{PR}, S\textsubscript{BS}, S\textsubscript{BV}) settings, reflecting a diverse range of contexts in which XR recording may occur. The scenarios were also evenly divided between situationally impaired and non-impaired contexts. Activities such as reading a book (S\textsubscript{PR}), filling out a form (S\textsubscript{BF}), taking an exam (S\textsubscript{CE}), and playing volleyball (S\textsubscript{BV}) may impair a bystander’s situational awareness due to visual or cognitive engagement. Note that this study only accounted for the presence of situational impairment; the degree or severity of impairment was not in the scope of this exploratory design work. Furthermore, the scenarios varied in privacy sensitivity. Some scenarios involved private activities such as filling out personal information (S\textsubscript{BF}), or reading sensitive material (S\textsubscript{PR}, S\textsubscript{CE}), while others reflected more casual or public settings, such as waiting in line (S\textsubscript{BW}), playing a game (S\textsubscript{BV}), or sitting on a park bench (S\textsubscript{PB}). This range of conditions enables a nuanced examination of how situational impairments and privacy contexts influence perceptions of XR recording and privacy indicators.

\textbf{System Usability Scale (SUS): } In each section, participants completed a SUS questionnaire, which uses a five-point scale ranging from ``strongly disagree'' to ``strongly agree''. The SUS results ensured that the proposed systems met acceptable usability standards. Since participants did not interact directly with the systems, the questionnaire was adapted to assess their expectations based on the descriptions provided. The specific questions were as follows:

\begin{enumerate}
    \item I think I would like to use this system frequently.
    \item I find this privacy indicator unnecessarily complex.
    \item I think this privacy indicator is easy to use.
    \item I think I would need technical support to use this privacy indicator.
    \item I find the functions in this privacy indicator well integrated.
    \item I think there is too much inconsistency in this privacy indicator.
    \item I would imagine that most people would learn to use this privacy indicator quickly.
    \item I find this privacy indicator cumbersome to use.
    \item I feel confident using this privacy indicator.
    \item I need to learn a lot before I can use this privacy indicator.
\end{enumerate}

\begin{figure*}[t!]
    \centering
    \includegraphics[width=0.9\linewidth]{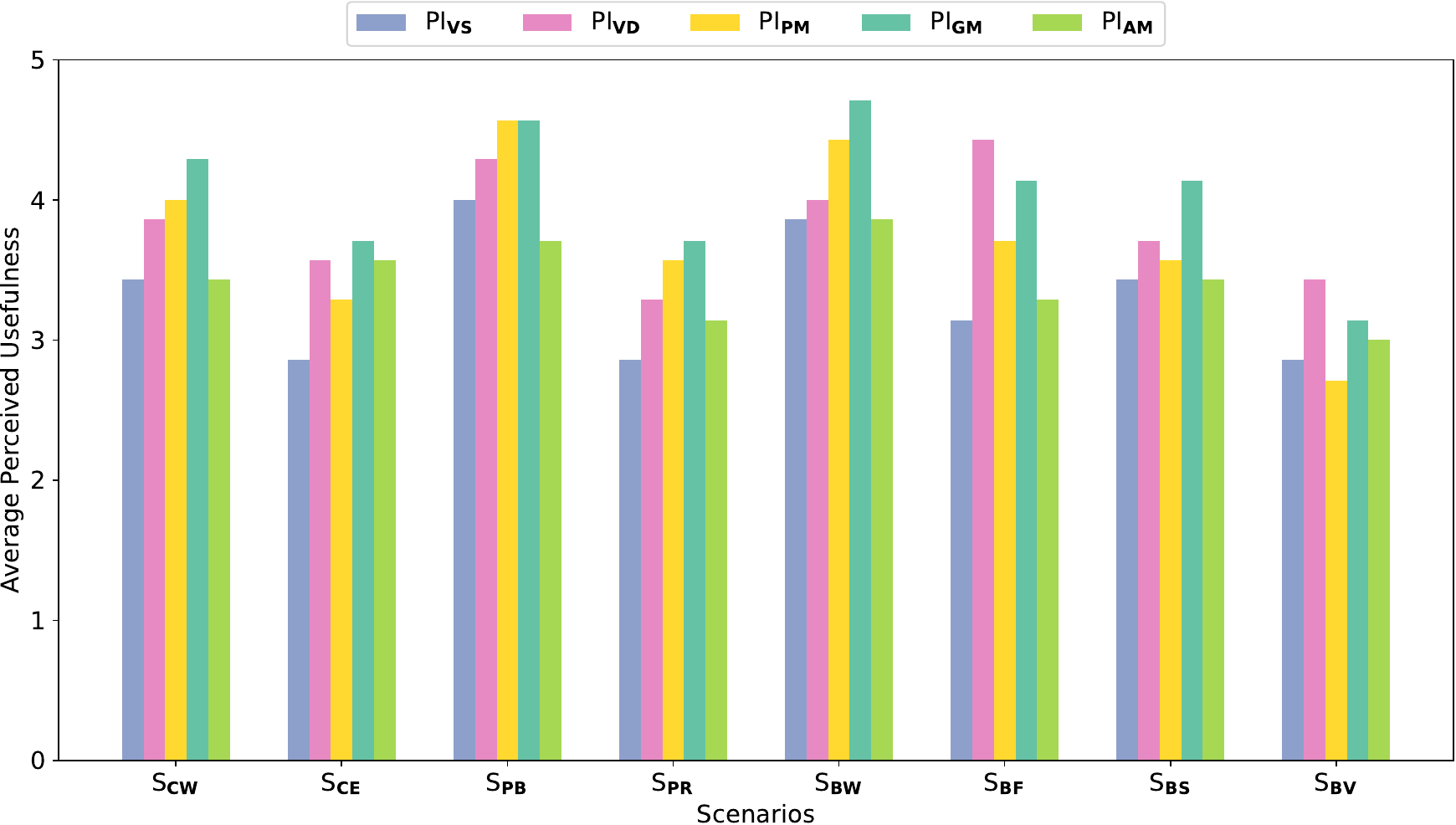}
    \caption{Perceived usefulness of proposed privacy indicators per scenario averaged across participants.
    }
    \label{fig:scenarios}
\end{figure*}

\begin{figure}[t!]
    \centering
    \includegraphics[width=\linewidth]{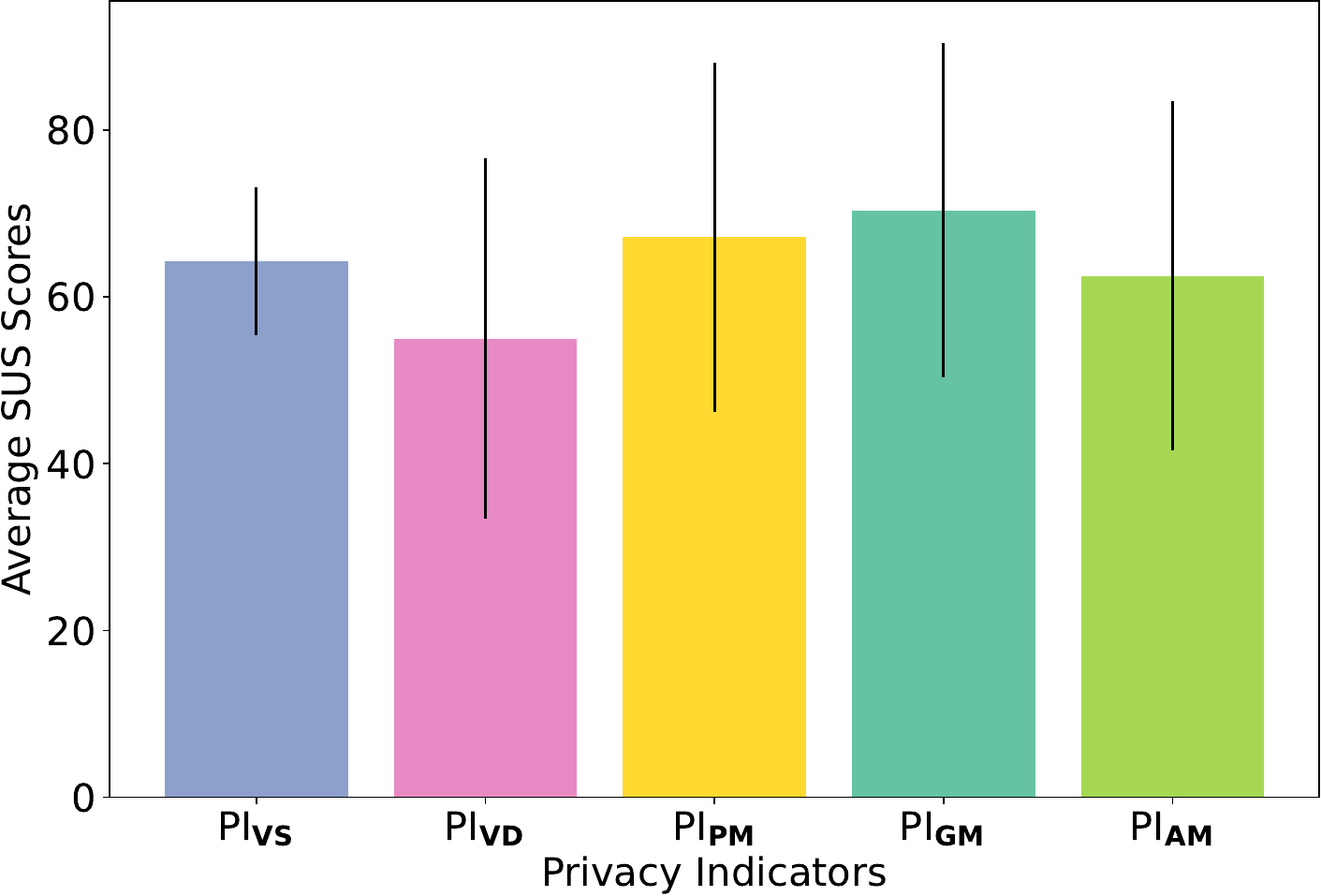}
    \caption{Average SUS scores for each proposed privacy indicator.}
    \label{fig:sus}
\end{figure}

\textbf{Preference Ranking: } In the final section of the survey, participants ranked the five proposed XR privacy indicators based on their preferences for general use, with five representing the highest ranking and one representing the lowest ranking. Participants then re-ranked the systems from the perspective of a situationally impaired bystander. After completing the rankings, participants were asked to justify their choices for the highest and lowest-ranked systems in both contexts.

Lastly, participants were given the opportunity to provide any additional thoughts or feedback through an open-ended question at the end of the survey.

\subsection{Results}\label{userstudy-results}

We report our findings from the user study in three parts. First, we assess the usability of each proposed privacy indicator using the System Usability Scale (SUS) and report the average SUS scores in~\S\ref{userstudy-results-systemsusabilityscores}. Next, we analyze the perceived usefulness of each indicator across different privacy and situational impairment scenarios in~\S\ref{userstudy-results-perceivedusefulness}. Finally, we present the preference rankings of the participants for the proposed indicators, comparing how these preferences shift between non-impaired and situationally impaired contexts in~\S\ref{userstudy-results-preferencerankings}.

\subsubsection{System Usability Scores}\label{userstudy-results-systemsusabilityscores}

We administered SUS to evaluate the usability of the proposed XR privacy indicators. As shown in Figure~\ref{fig:sus}, all five indicators received average SUS scores above the accepted usability threshold of 50~\cite{bangor2009determining}. PI\textsubscript{GM} and PI\textsubscript{PM} achieved the highest usability ratings, while PI\textsubscript{VD} scored the lowest. These results validate that each design meets a basic standard of usability.

\subsubsection{Perceived Usefulness}\label{userstudy-results-perceivedusefulness}

We also measured the perceived usefulness of each proposed privacy indicator across a range of bystander scenarios differing in privacy sensitivity and situational impairment. As shown in Figure~\ref{fig:scenarios}, participants rated indicators as more useful in scenarios where the bystanders were not situationally impaired and privacy concerns were minimal (e.g., S\textsubscript{CW}, S\textsubscript{PB}, and S\textsubscript{BW}). Conversely, the perceived usefulness ratings declined in situationally impaired, privacy-sensitive scenarios (e.g., S\textsubscript{CE}, S\textsubscript{BF}, and S\textsubscript{BV}). This trend suggests that participants were more critical of the effectiveness of privacy indicators when their privacy expectations were high and attention was limited. Moreover, PI\textsubscript{GM} was ranked as the most useful indicator in six of the eight scenarios. This trend reinforces the robustness of PI\textsubscript{GM} and aligns with its strong usability score shown in Figure~\ref{fig:sus}. Furthermore, \emph{indicators based solely on visual signals (PI\textsubscript{VS}, and PI\textsubscript{VD}) were poorly rated for perceived usefulness}.

\subsubsection{Preference Rankings}\label{userstudy-results-preferencerankings}

\begin{figure}[t!]
    \centering
    \includegraphics[width=\linewidth]{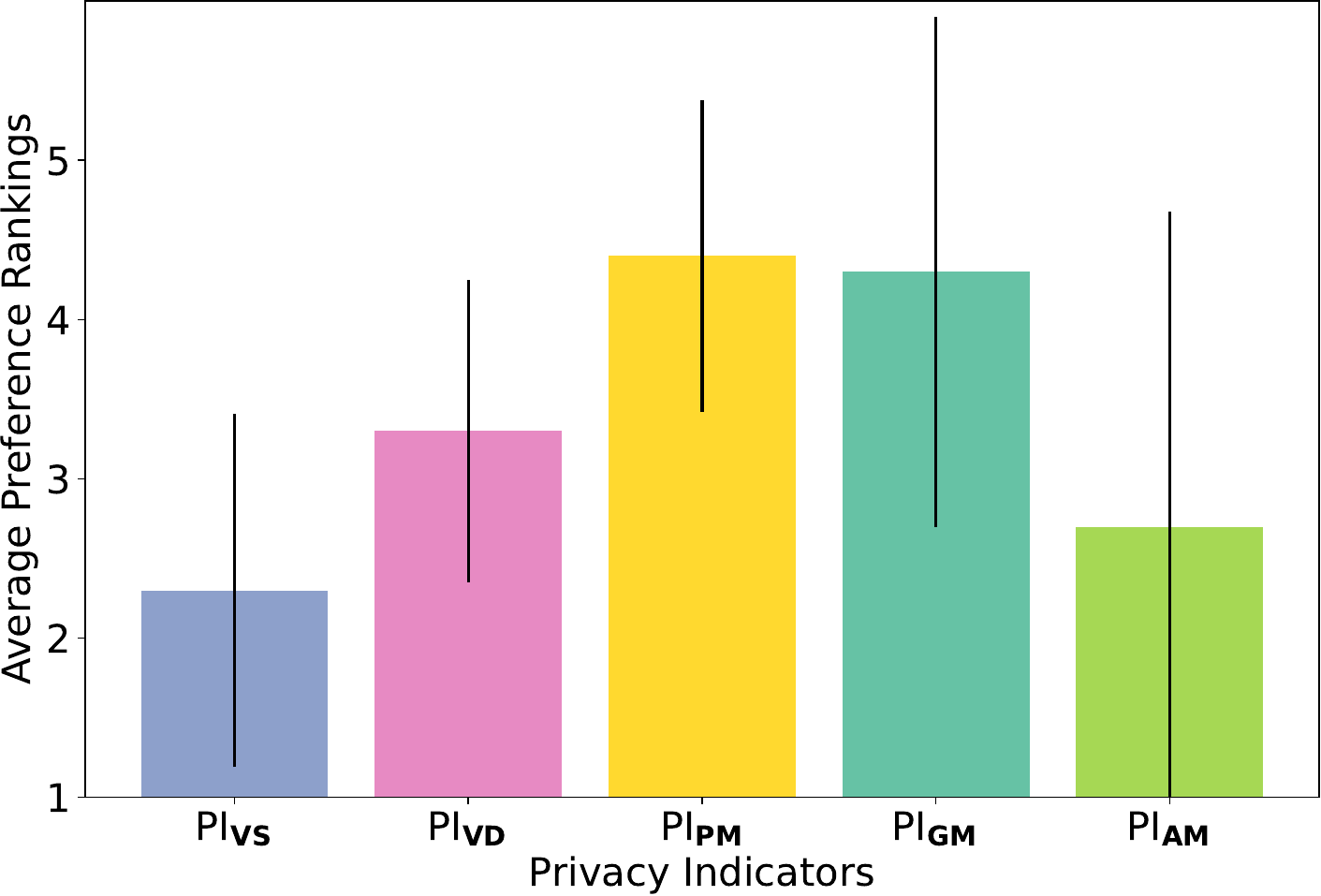}
    \caption{Average preference rankings of proposed privacy indicators for non-situationally impaired bystanders.}
    \label{fig:ranks-non-si}
\end{figure}

\begin{figure}[t!]
    \centering
    \includegraphics[width=\linewidth]{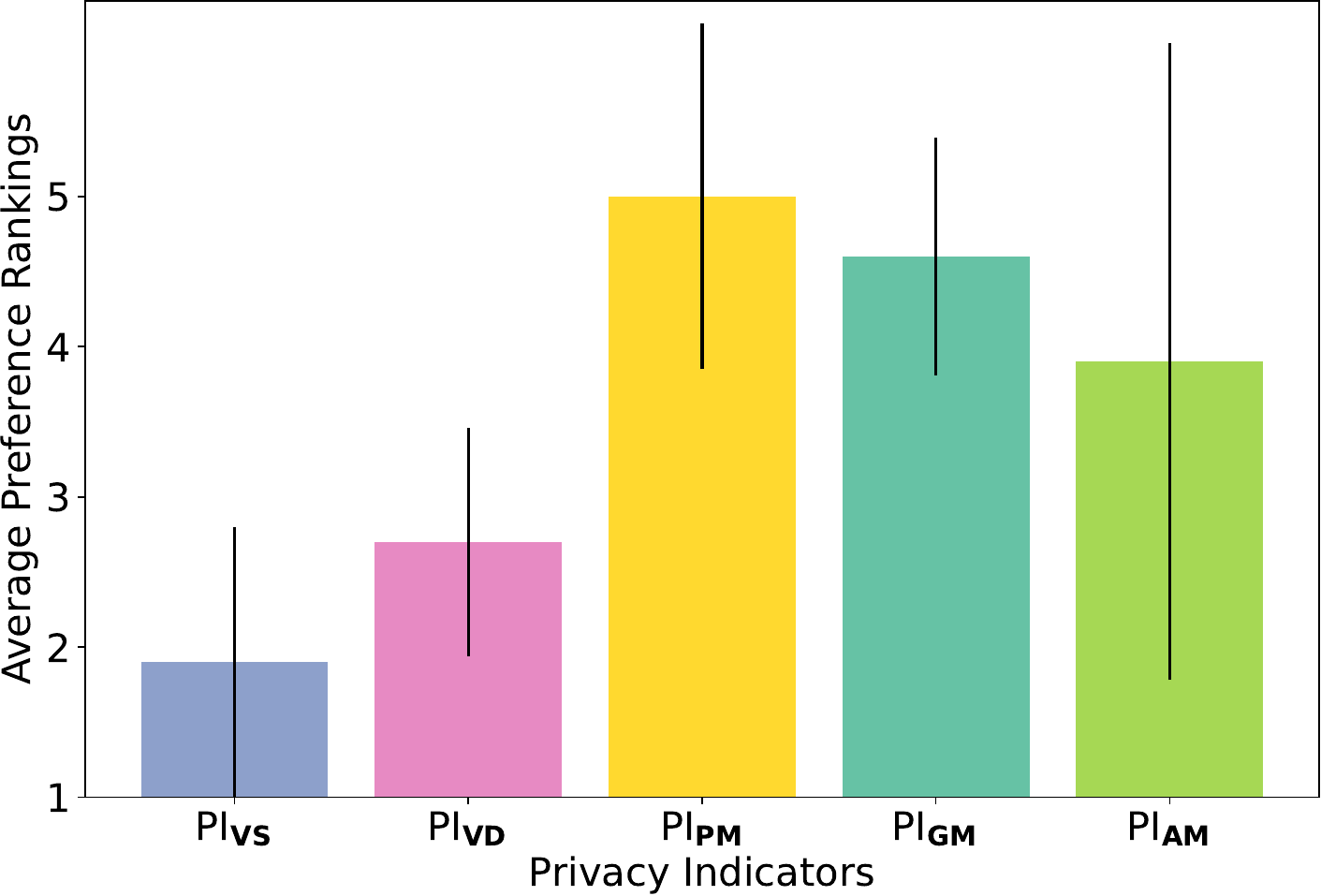}
    \caption{Average preference rankings of proposed privacy indicators for situationally impaired bystanders.}
    \label{fig:ranks-si}
\end{figure}

To explore the subjective preferences for the proposed privacy indicators, participants ranked the indicators from most preferred (6) to least preferred (1), as both non-situationally impaired (Figure~\ref{fig:ranks-non-si}) and situationally impaired (Figure~\ref{fig:ranks-si}) bystanders.


Participants ranked PI\textsubscript{AM} second worst, on average, for non-situationally impaired bystanders, citing its perceived intrusiveness. For example, \textit{P32} described it as causing \textit{ ``unnecessary disturbance''}. However, the ranking saw an increase, from $2.7 \pm 1.9$ to $3.9 \pm 2.1$, when bystanders became situationally impaired. This trend highlights a shift in perceived value. Participants appreciated the multi-sensory cues offered by PI\textsubscript{AM} (flashing light, escalating sound), since they were perceived as hard to miss, even in distracting environments. For example, \textit{P23} remarked that PI\textsubscript{AM} \textit{``would redirect my attention most easily''}.

On the other hand, PI\textsubscript{VS} was consistently ranked the worst by bystanders with and without situational impairment. Participants described it as too subtle and ineffective at a distance or when not directly in view. For example, \textit{P06} stated that PI\textsubscript{VS} was \textit{``not easy to notice if I am occupied''}. Some participants also highlighted the lack of information conveyed by such indicators. For instance, \textit{P25} pointed out that PI\textsubscript{VS} \textit{``only tells me when someone is recording, not what or who or for how long''}. This result highlights the need to move beyond conventional designs and develop privacy indicators that are inclusive for situationally impaired bystanders.

\section{Discussion}\label{discussion}

In this section, we reflect on the broader implications of our findings and discuss key design recommendations for developing inclusive XR privacy indicators~(\S\ref{discussion-designrecs}). We also examine practical considerations for real-world deployment of the proposed indicators~(\S\ref{discussion-deploymentconsiderations}), discuss the ethical and user experience trade-off~(\S\ref{discussion-ethicalanduxtradeoffs}), and outline our study's limitations alongside directions for future work~(\S\ref{discussion-limitationsandfuturework}).

\subsection{Design Recommendations}\label{discussion-designrecs}

Our findings reinforce the central motivation of this work: the need to design XR privacy indicators that remain effective for situationally impaired bystanders. The user study participants were consistently more critical of the usefulness of proposed indicators in scenarios where both privacy sensitivity and situational impairment were present (see \S\ref{userstudy-results-perceivedusefulness}). This result clearly highlights that traditional, one-size-fits-all privacy indicators that assume constant visibility and attention from bystanders fail to meet their privacy demands in real-world contexts.

A consistent trend across both perceived usefulness (\S\ref{userstudy-results-perceivedusefulness}) and preference rankings (\S\ref{userstudy-results-preferencerankings}) was the \emph{poor performance of visual-only indicators}. Its resemblance to current commercial implementations (e.g., LED-based indicators) highlights the limitations of relying solely on visual signals. Such indicators were frequently described as too subtle, easily missed when not directly in view, or uninformative without added context by the user study participants. The fact that PI\textsubscript{VS} performed poorly despite its familiarity highlights a key design pitfall: familiarity does not guarantee effectiveness when it comes to designing XR privacy indicators, especially for distracted bystanders. These findings reinforce the concerns raised in prior work about the inadequacy of such indicators for situationally impaired bystanders.

In contrast, our proposed \emph{indicators with multimodal cues were consistently rated higher in both usability (\S\ref{userstudy-results-systemsusabilityscores}) and perceived usefulness}, particularly in scenarios with situationally impaired bystanders. These results suggest that inclusive XR privacy indicator design must explore modality diversity to remain perceptible and informative under situational impairment, ensuring bystanders’ privacy is respected even when their attention or awareness is limited. In addition to modality diversity, these results highlight that inclusive indicators must adapt to context.

Collectively, our findings highlight key design principles for developing inclusive XR privacy indicators. Designers should avoid over-relying on static visual cues, as they often assume ideal viewing conditions are present. Instead, multimodal indicators, incorporating visual, auditory, or haptic signals, should be preferred as they can enhance noticeability across diverse contexts. Indicators should also be designed to be context-aware, adapting their salience or modality based on contextual factors such as the activity being performed (e.g., passive sensing vs. active recording) and environmental factors (e.g., crowded vs. quiet spaces). Furthermore, inclusive indicators should not only make the bystanders aware of their privacy but also aim to provide them with intelligible information. For example, who is recording them and for how long? These insights offer practical design directions that facilitate the design of inclusive privacy indicators for the nuanced realities of real-world bystander experiences in XR environments.

\subsection{Deployment Considerations}\label{discussion-deploymentconsiderations}

While our study is focused on the design and evaluation of novel XR privacy indicators to support situationally impaired bystanders, the real-world deployment of the proposed indicators presents several technical challenges. The proposed multimodal indicators require hardware capabilities that are not yet uniformly available across XR devices. For example, gaze-based notification systems depend on reliable eye tracking, which is only present in higher-end XR devices. Similarly, smartphone-based alerts require coordination across device ecosystems and assume that bystanders have compatible devices and permissions enabled. Energy consumption is another practical constraint, particularly for persistent sensing or real-time face detection. Deploying adaptive or escalating indicators may require continuous monitoring of user activity and surroundings, which can raise system complexity and result in battery drainage. These challenges highlight the importance of considering platform-specific constraints and optimization for efficiency and interoperability in future implementations.

\subsection{Balancing Noticeability and Social Acceptability}\label{discussion-ethicalanduxtradeoffs}

Our results highlight a trade-off between the noticeability of privacy indicators and potential social disruption. While participants favored subtle indicators in low-privacy or non-impaired scenarios, they expressed appreciation for more salient and disruptive cues when situational impairment was introduced. This trend presents a key design trade-off: sufficiently noticeable privacy indicators may also risk drawing unwanted attention or causing discomfort in social settings. Cultural norms and environmental factors may further modulate this balance. For example, loud audio cues may be inappropriate in libraries, classrooms, or certain public spaces, while overt visual displays could be misinterpreted in different cultural contexts. Future work should explore adaptive systems that tailor the modality and intensity of privacy indicators based on situational and cultural context, supporting effectiveness and acceptability.

\subsection{Limitations and Future Work}\label{discussion-limitationsandfuturework}

A key limitation of our work is that user study participants evaluated proposed privacy indicators based on their written descriptions, rather than interacting with their functioning prototypes. We attempted to mitigate this limitation by providing detailed descriptions of each indicator and allowing participants to ask clarification questions to ensure a shared understanding. However, their evaluations may still not fully reflect how these indicators would perform in real-world use. In future work, we plan to leverage generative AI to create visual and interactive prototypes of indicators to enable more realistic and embodied evaluations of the proposed indicators.

Moreover, we treated situational impairment as binary (bystanders were assumed to be either impaired or not) in our scenarios for the user study. However, the real-world attention and perceptual capacity of bystanders are more complicated than that. For example, someone navigating a crowded street while engaged in conversation may be less able to perceive subtle cues (more situationally impaired) than someone momentarily distracted (less situationally impaired). In future work, we plan to develop scenarios that span a range of situational impairment of bystanders, informed by behavioral or physiological data.

We also note that our study focused primarily on individual bystanders, yet XR systems increasingly operate in multi-bystander, socially dynamic environments such as classrooms, transit hubs, or collaborative workspaces. We acknowledge that the indicators that work well in isolation may behave differently in collective contexts, where privacy norms are co-constructed. In future work, we plan to explore broader social and environmental dynamics, including multi-bystander scenarios. By leveraging interactive prototypes of our designs, we also aim to conduct longitudinal studies that capture how perceptions of indicator effectiveness evolve with repeated exposure, shifting social contexts, and increased familiarity.

Another limitation of this work is the small sample size of the user study. An a priori power analysis indicated that 28 participants would be required to achieve the desired statistical power for our repeated-measures Analysis of Variance (ANOVA)
design. However, due to the exploratory nature of this study and resource constraints, we recruited only seven
participants. We did not conduct statistical significance testing, and the patterns and differences observed should therefore be interpreted as preliminary trends rather than statistically validated effects. This limits the generalizability of the findings. Future work will include a larger-scale study to validate the results and confirm the observed patterns with a sufficiently powered participant pool.

\section{Conclusion}\label{conclusion}
This work demonstrates that current XR privacy indicators fall short in supporting situationally impaired bystanders. Our participatory design sessions and user evaluations show that visual-only indicators often fail, especially in privacy-sensitive scenarios where bystanders experience reduced attention or awareness. We found that multimodal indicators, such as combining visual, auditory, and haptic cues, provide more inclusive and effective signaling. These findings emphasize the importance of designing XR privacy indicators that offer sensory diversity and adapt to contextual and situational demands. Designers must build systems that ensure bystanders can perceive, interpret, and respond to privacy cues, regardless of their momentary impairments. By adopting adaptable, multimodal, and context-aware approaches, future XR systems can promote more trustworthy and equitable interactions in shared environments.

\acknowledgments{
The authors acknowledge support from the Commonwealth Cyber Initiative, the National Science Foundation under Award No. CNS-2350116, and the United States Military Academy under Cooperative Agreement No. W911NF2520133. Any opinions, findings, and conclusions or recommendations expressed in this material are those of the author(s) and do not necessarily reflect the views of the Commonwealth Cyber Initiative, the National Science Foundation, or the United States Military Academy. The authors also thank Ahmad Faraz Khan and Tianyang Robin Lu for their valuable contributions during the early stages of this project. 

Moreover, the authors would like to acknowledge the use of ChatGPT4 to generate the teaser image and to revise the text throughout all sections of the paper to correct typos, grammatical errors, and awkward phrasing.}

\bibliographystyle{abbrv-doi}

\bibliography{template}

\end{document}